\documentclass[a4paper]{jpconf}

\usepackage{amscd,amsmath}

\newcommand\ointint{\begingroup 
\displaystyle \unitlength 1pt 
\int\mkern-7.2mu 
\begin{picture}(0,3) 
\put(0,3){\oval(10,8)} 
\end{picture} 
\mkern-7mu\int\endgroup} 

\begin{document}

\title{The emergence of gravity as a retro-causal post-inflation macro-quantum-coherent holographic vacuum Higgs-Goldstone field}

\author{Jack Sarfatti$^1$ and Creon Levit$^2$}
\address{$^1$ Internet Science Education Project}
\address{$^2$ NASA Ames Research Center}
\ead{adastra1@mac.com  creon.levit@nasa.gov }

\begin{abstract}

 We present a model for the origin of gravity, dark energy and dark
 matter: Dark energy and dark matter are residual pre-inflation false
 vacuum random zero point energy ($w\!=\!-1$) of large-scale negative,
 and short-scale positive pressure, respectively, corresponding to the
 ``zero point'' (incoherent) component of a superfluid (supersolid)
 ground state. Gravity, in contrast, arises from the 2nd order
 topological defects in the post-inflation virtual ``condensate''
 (coherent) component.  We predict, as a consequence, that the LHC
 will never detect exotic real on-mass-shell particles that can
 explain dark matter $\Omega_{\mathrm{DM}} \approx 0.23$. We also
 point out that the future holographic dark energy de Sitter horizon
 is a total absorber (in the sense of retro-causal Wheeler-Feynman
 action-at-a-distance electrodynamics) because it is an infinite
 redshift surface for static detectors. Therefore, the advanced
 Hawking-Unruh thermal radiation from the future de Sitter horizon is
 a candidate for the negative pressure dark vacuum energy.
\end{abstract}

\section{Gravity from topological singularities in the quantum vacuum}

We consider the possibility that the Einstein-Cartan 1-forms
consistent with 1915 General Relativity (GR) are local macro-quantum
emergent {\em supersolid} \cite{Sarfatti} c-number fields.  We mean
this in the same sense that $v$ (the locally irrotational superflow 3D
Galilean relativity group velocity 1-form in superfluid
\textsuperscript{4}He) is emergent, with quantized
circulation.  The single-valuedness of the associated $S^1$
macroquantum coherent Higgs-Goldstone order parameter
$\Psi=|\Psi|e^{i\Theta}$ emerges from an effective spontaneous broken \cite{Stoneham}
non-electromagnetic $U(1) = O(2)$ ground state gauge symmetry\footnote{This corresponds to Hagen Kleinert's
  multi-valued singular phase transformations (discussed elsewhere in
  these proceedings).
}.

\begin{equation}
v= \frac{1}{2\pi }\frac{h}{m}\mathrm{d\Theta }
\end{equation}
\begin{equation}
\oint v=\frac{1}{2\pi }\frac{h}{m}\oint d\Theta =n\frac{h}{m}\\
\end{equation}
\begin{equation}
n = \pm 1,\pm 2, ...
\end{equation}

In analogy to the above, we use a phenomenological model for the
moment of inflation with eight macroquantum coherent relative phase
0-forms $\Theta^I$ and $\Phi^I$ that form two Lorentz group 4-vectors
with magnitudes $\Theta$ and $\Phi$, respectively. These magnitudes,
in turn, are the phases of a $S^2$ vacuum order parameter
manifold. This $S^2$ fiber bundle over real spacetime supports stable
point monopole topological defects (simultaneous nodes of the three
corresponding Higgs fields) \cite{Thouless}.  

These GeoMetroDynamic (GMD) point monopoles are the lattice points in
spacelike slices of Hagen Kleinert's ``world crystal lattice''
\cite{Kleinert}.  They correspond to a non-trivial
2{}\textsuperscript{nd} homotopy group of emergent effective
post-inflation field $O(3)$ mappings of surrounding non-bounding
2-cycles S{}\textsuperscript{2}${}_{3D}$ in 3D physical space to the
vacuum manifold S{}\textsuperscript{2}.

The inhomogeneities in all eight phases $\Theta^I$ and $\Phi^J$ form the
emergent GMD tetrad field $A^I$.  For
details, see equations 14-20, below.

In the world hologram conjecture \cite{Susskind}\cite{Hogan}, with total hologram
screen area A, the mean separation $\Delta$L of the lattice points is
for our pocket universe c/H${}_{0} = \sqrt{N}L_P = 10^{29}$ cm on the
cosmic landscape of the megaverse, given by\cite{Ng}:

\begin{equation}
\Delta L=\left( {\mathrm{L{}}}_{P}{}^{2}\sqrt{A}\right) {}^{
1/3} = \left( {\mathrm{L{}}}_{P}{}^{2}\frac{c}{{\mathrm{H{}}}_{0}}\right)
{}^{1/3}\ \approx 10{}^{-13}\mathrm{cm}\\
\end{equation}
\begin{equation}
N=\frac{A}{{4\mathrm{L{}}}_{P}{}^{2}}
=\frac{1}{4\Lambda {{\mathrm{L{}}}_{P}{}^{2}}}
=\frac{A^{\left(
3/2\right) }}{\left( \Delta L\right) {}^{3}}=\frac{V}{\left( \Delta
L\right) {}^{3}}\\
\end{equation}
\begin{equation}
A = \partial V
\end{equation}

where $ \partial $ is the quasi-boundary operator because the
surrounding future light cone surface is a non-bounding de Rham
2-cycle.  This 2-cycle encloses $N$ GMD point monopole defects in the
three effective real $O(3)$ Higgs field macroquantum coherent
Penrose-Onsager off-diagonal-long-range-order post-inflation vacuum
parameters\cite{Yang}.

Wrapping once around S{}\textsuperscript{2}${}_{3D}$ through solid
angle 4$\pi$ wraps an integer $N$ times round the vacuum manifold
S{}\textsuperscript{2}. This is analogous to the global quantized
circulation vortices in superfluid \textsuperscript{4}He that are the
stable topological defects in the first homotopy group for an $O(2)$
mapping with only a single relative Goldstone phase $\Theta$ and two
real Higgs scalars $\Psi$${}_{1}$, $\Psi$${}_{2}$ (instead of the
three $\Psi$${}_{1}$, $\Psi$${}_{2}$,$\Psi$${}_{3}$ with two relative
Goldstone phases $\Theta$, $\Phi$ over 3D spacelike slices of physical
spacetime in our toy model). 

This wrapping integer $N$ is the explanation for the Bekenstein bit
quantized areas of null black hole event horizons and
observer-dependent cosmological (e.g. dark energy future de Sitter)
horizons\footnote{In this paper we use the term ``de Sitter horizon''
  informally.  A more precise description would be ``future
  cosmological event horizon''.}  of area A that obey Hawking's
entropy and temperature formulas
\begin{equation}
S={\mathrm{k{}}}_{B}A/4{\mathrm{L{}}}_{P}{}^{2} = {\mathrm{Nk{}}}_{B}\\
\end{equation}
\begin{equation}
T=\frac{\partial E}{\partial S} = \frac{\mathrm{hc}}{{\mathrm{k{}}}_{B}\sqrt{A}} 
= \frac{\mathrm{hc}\sqrt{\Lambda}}{{\mathrm{k{}}}_{B}}
= \frac{\mathrm{hc}}{{\mathrm{k{}}}_{B}{\mathrm{L{}}}_{P}\sqrt{N}}
\end{equation}
\begin{equation}
\ointint 2d\Theta \wedge d\Phi = 4 \pi N =
A/4{\mathrm{L{}}}_{P}{}^{2} = 10{}^{124} \ \mathrm{Bekenstein} \  \mathrm{BITs,}
\end{equation}
where the double integral around the vacuum manifold is induced by a
single wrap around the future asymptotic
de Sitter horizon. The de Sitter horizon is a surrounding
(but non-bounding \cite{Kiehn}) closed 2-cycle at lightlike conformal
infinity. It is also a stretched thermal horizon for comoving observers
in the accelerating Hubble flow \cite{Susskind}. 

The remaining six Goldstone phase angles form the Calabi-Yau space of
string theory - the same field Gennady Shipov calls the ``oriented
point'' \cite{Shipov}.  Einstein's 1915 curvature field is simply the local gauge
field from the 4-parameter translation universal spacetime symmetry
group $T(4)$ for all matter fields (i.e., strong equivalence principle)
with the constraint of zero torsion. 

Locally gauging the 10-parameter Poincare group $P(10)$ of Einstein's
1905 special relativity gives the Einstein-Cartan theory of ({\em
  dislocation} defect) torsion\cite{Kleinert} in addition to ({\em
  disclination} defect) curvature\footnote {There are two classes of
  defects: The monopole defects which form the ``atoms'' of the
  supersolid world crystal lattice, and the disclinations and
  dislocations in this lattice, which account for, respectively, the
  curvature and torsion of spacetime.}  of the symmetric Levi-Civita
connection. Indeed, this local gauge field model can be reinterpreted
in terms of the eight multi-valued Goldstone phases of the coherent
post-inflation vacuum field. The Calabi-Yau space seems to be simply
the torsion field in disguise.

\section{Dark energy from the future}

The future de Sitter event horizon world hologram is ``our past light
cone at the end of time''\cite{Davis}.  It can be pictured as a
pixelated spherical shell of area NL${}_{P}${}\textsuperscript{2}
infinitely far from our detectors (in proper time) on their {\itshape
  future} light cone, with thickness L${}_{P}$ and duration
L${}_{P}$/c. This shell, or ``screen'', has 4D volume
NL${}_{P}${}\textsuperscript{4} with dark energy density hc/(4DVolume
Hologram Screen). This screen projects the voxels of our accelerating
expanding 3D space hologram image back from the future - indeed, back
to the moment of inflation 13.7 billion years ago in what Igor Novikov
calls a ``globally self-consistent'' strange loop in time.

To summarize: The area of an observer's future de Sitter horizon
holographically determines the dark energy density seen by that observer.

For a static local-non-inertial-frame (LNIF) observer (with covariant
acceleration $g = c^{2}/\sqrt{N}L_{P} = cH_0 \approx 10^{-9}$
m/sec\textsuperscript{2} relative to the $\Lambda\!=\!0$, $k\!=\!0$
spatially flat post-inflation background Friedman metric) the
Hawking-Unruh temperature of the future de Sitter horizon is
proportional to his acceleration.  This is in similar to a static
outside observer adiabatically approaching the horizon of a black hole
-- being ``slowly lowered down on a cable'' -- who measures a
temperature which approaches the Planck temperature
hc/L${}_{P}$k${}_{B}$ as he approaches the horizon.

Of course, the locally coincident geodesic observer relative to the
Friedman metric sees no heat radiation - only $w=-1$ positive dark
zero point energy density vibrations of equal but opposite negative
pressure per large space dimension, as required by the Einstein
Equivalence Principle (EEP). 

In contrast the NASA WMAP isotropic black body radiation in the
comoving Friedman frame is coming along our past light cone from the
surface of last contact 380,000 years after the post-inflation
reheating of the Big Bang, and its energy density weakens $a(t)^{-4}$ as
space expands because it has $w = +1/3$ ratio of pressure to energy
density\footnote {There is a lack of consensus and clarity in the
  literature on who sees what. The Unruh effect in globally flat
  Minkowski spacetime is: A covariantly proper accelerating local
  detector (not on a timelike geodesic, which by definition has zero
  covariant proper 4-acceleration) sees thermal equilibrium blackbody
  radiation whose temperature is proportional to its covariant
  4-acceleration magnitude. In contrast, a momentarily coincident
  non-accelerating detector sees only zero point vacuum fluctuations
  instead of the thermal radiation. That is, some of the vacuum
  fluctuation energy is converted into thermal radiation in the rest
  frame of the intrinsically accelerating detector. 

  Static detectors outside the event horizon of an ideal Schwarzschild
  black hole are covariantly properly accelerating in order to ``stand
  still'' at a fixed Schwarzschild radial coordinate $r$. This is in
  accord with the actual Pound-Rebka Harvard Tower experiment showing
  the gravity redshift using the nuclear Mossbauer
  effect. Furthermore, the comoving detectors in the Robertson-Walker
  representation at constant $\chi$ are analogous to the previous case
  at constant $r$ \cite{Davis}:

\[
ds^2 = -c^2dt^2+R(t)^2[d\chi^2+S_k^2(\chi)d\psi^2],
\]

  However, e.g., Davies \& Davis \cite{Davies} write ``the response of a
  particle detector travelling along a geodesic in a de Sitter
  invariant vacuum state; the detector behaves as if immersed in a
  bath of thermal radiation'' Of course,``geodesic'' depends on choice
  of the local GCT frame invariant gravity tetrad field. Thus, the
  detector on a geodesic in the de Sitter gravity tetrad field is
  actually properly accelerating with respect to the geodesic in the
  zero cosmological constant Robertson-Walker gravity tetrad
  field. This is the point of view we take in this paper and it agrees
  operationally with how the dark energy data is actually interpreted.
}.

We propose that the dark energy zero point vacuum fluctuations
measured by non-rotating covariantly unaccelerated Local Inertial
Frame (LIF) detectors (on timelike geodesics relative to the physical
spacetime multi-valued connection local gauge field
$\Gamma${}\textsuperscript{$\mu$}${}_{\nu \lambda }$ that forms curved
and torsioned spacetime) appear as advanced Wheeler-Feynman
quasi-thermal blackbody ``Unruh radiation''.  It comes back from the
future de Sitter horizon ``perfect absorber'' with temperature that
has order of magnitude

\begin{equation}
T = \frac{\mathrm{hg}}{{\mathrm{ck{}}}_{B}}\rightarrow \frac{\mathrm{hc}}{{\mathrm{k{}}}_{B}{\mathrm{L{}}}_{P}\sqrt{N}} \approx \frac{\mathrm{hc}}{{\mathrm{k{}}}_{B}{\mathrm{L{}}}_{P}10{}^{62}}
\end{equation}
for covariantly accelerated Local Non-Inertial Frame (LNIF) detectors
off timelike geodesics.

For example, a static observer outside the event horizon of a
non-rotating black hole must covariantly accelerate away from
the black hole radially with 

\begin{equation}
g = -\frac{GM}{r^2} \frac{1}{\sqrt{1-\frac{2GM}{c^2r}}} \sim T
\end{equation}
in order to stay at fixed r in the curved spacetime outside the black
hole. They need to fire their rocket engines in order to remain
static. These static LNIF observers see the event horizon as a
``stretched membrane'' with Unruh temperature $T$. Coincident LIF
observers do not see this at all. This is an example of what Leonard
Susskind calls ``horizon complementarity'' \cite{Susskind}, in analogy
with Niels Bohr's quantum complementarity of wave-particle duality
from the non-commutativity of the Lie algebra of observable operators
on qubit Hilbert space fibers over classical field configuration
space.

The comoving observers that see an approximately isotropic WMAP Cosmic
Microwave Background (CMB) in our accelerating expanding universe are
analogous to the static LNIF observers in the Schwarzschild model
where now there is a universal acceleration $g \approx 10^{-9}$ m/sec{}\textsuperscript{2},
which is the same order of magnitude of the anomalous Sun-centered
radial accelerations of the two NASA Pioneer space probes beyond the
orbit of Jupiter. This is a curious coincidence that possibly has
deeper significance, although it is surprising to find Hubble's
parameter $H_0$ appearing on such a short scale local metric field. In
contrast LIF detectors see this advanced quasi-thermal Unruh
Wheeler-Feynman radiation as zero point vacuum fluctuation energy
density

\begin{equation}
  \rho {}_{\mathrm{DE} } = {\mathrm{string\ tension}}\times{\mathrm{vacuum\ curvature}}
  = \frac{\mathrm{string\ tension}}{\mathrm{area\ of\ future\ cosmic\ horizon}}
\end{equation}
\begin{equation*}
  = \frac{\mathrm{hc}}{{\mathrm{NL{}}}_{P}{}^{4}} =
\frac{\mathrm{hc}}{\left( 10{}^{-2}\mathrm{cm}\right) {}^{4}}=  0.73\times{10}^{-29}\mathrm{grams}/\mathrm{cc}.
\end{equation*}

It is as if there is an effective high frequency cutoff at
$c/10{}^{-2}\mathrm{cm} = 3\times 10^{12}$ Hz for the $w=-1$ zero
point dark energy virtual photon vibrations with critical wavelength
equal to the geometric mean of the future de Sitter horizon scale and
the Planck scale.  The world hologram model posits that the number N
of interior 3D voxels N of size $\Delta$L {\em equals} the number of
2D pixels of size L${}_{P}$ on the world hologram future de Sitter
horizon.

\section{Calabi-Yau from torsion. Brane theory from the 1970s}

The four tetrad 1-form fields e{}\textsuperscript{I} are the General
Coordinate Invariant (GCI) gravitational fields in Einstein's 1915
GR. They form a single 4-vector under the 6-parameter homogeneous
Lorentz group $SO(1,3)$. The non-trivial curvilinear 4D General Coordinate
Transformations (GCT) connect covariantly accelerating coincident
LNIFs with g-forces on its rest detectors. A non-gravity force is
required to create a translational covariant acceleration. Conservation
of angular momentum maintains a rotating LNIF in the absence of
friction in deep space once the external torque is removed. The
Lorentz group transformations connect coincident covariantly non-accelerating
LIFs with vanishing g-forces. The tetrad field components e{}\textsuperscript{$\mu$}${}_{I}$
and their inverses e{}\textsuperscript{I}${}_{\mu }$ connect locally
coincident LIFs with LNIFs. The LNIF curvilinear metric field is
g${}_{\mu \nu }$. The coincident LIF Center Of Mass (COM) metric
$\eta$${}_{{}_{\mathrm{IJ}}}$ is that of Minkowski space-time of
Einstein's 1905 SR. The Strong Equivalence Principle (SEP) implies
for the absolute differential local frame invariant ds
\begin{equation}
{\mathrm{ds{}}}^{2}= g{}_{\mu \nu }( \mathrm{LNIF}) {\mathrm{dx{}}}^{\mu
} {\mathrm{dx{}}}^{\nu } = \eta {}_{{}_{\mathrm{IJ}}}( \mathrm{LIF})
{\mathrm{e{}}}^{I}{\mathrm{e{}}}^{J}
\end{equation}

The multi-valued Goldstone phase transformations in our toy model
form a 4x4 M-Matrix of non-closed 1-forms where the non-trivial
parts of the four curvature-only tetrad 1-forms $A^I$
and the six non-trivial torsion field spin connection 1-forms $\varpi${}\textsuperscript{IJ}
= -$\varpi${}\textsuperscript{JI} are the diagonals and antisymmetrized
off-diagonal M-Matrix elements. 
\begin{equation}
{\mathrm{M{}}}^{\mathrm{IJ} }={\mathrm{d\Theta {}}}^{I}\wedge {\Phi
{}}^{J} - {\Theta {}}^{I} \wedge {\mathrm{d\Phi {}}}^{J} 
\end{equation}
\begin{equation}
{\mathrm{dM{}}}^{\mathrm{IJ} }= -2{\mathrm{d\Theta {}}}^{I}\wedge
{\mathrm{d\Phi {}}}^{J} 
\end{equation}
\begin{equation}
{\mathrm{d{}}}^{2} = 0
\end{equation}
\begin{equation}
{\mathrm{A{}}}^{I}= diag({\mathrm{M{}}}^{\mathrm{IJ}})
\end{equation}
\begin{equation}
{\mathrm{e{}}}^{I} = {\mathrm{I{}}}^{I} + {\mathrm{A{}}}^{I} =
{\mathrm{e{}}}^{I}{}_{\mu }{\mathrm{e{}}}^{\mu }{}_{\mathrm{LNIF}}
= {\mathrm{e{}}}^{\mu }{}_{I}{\mathrm{e{}}}^{I}{}_{\mathrm{LIF}}
\end{equation}
\begin{equation}
{\mathrm{I{}}}^{\mu }{}_{I} = {\delta {}}^{\mu }{}_{I}
\end{equation}
\begin{equation}
{\varpi {}}^{\mathrm{IJ}} = {\mathrm{M{}}}^{\left[ I,J\right] }
\end{equation}

The Einstein 1915 zero torsion field i.e. ${\varpi {}}^{\mathrm{IJ}}
$= 0,\ \ curvature field 2-form is 
\begin{equation}
{\mathrm{R{}}}^{\mathrm{IJ}} = {\mathrm{D\omega {}}}^{\mathrm{IJ}}
= {\mathrm{d\omega {}}}^{\mathrm{IJ}} + {\omega {}}^{I}{}_{K}\wedge
{\omega {}}^{K}{}^{J}\\
\end{equation}

Where the torsion field 2-form in Einstein-Cartan theory beyond
1915 GR would be
\begin{equation}
{\varpi {}}^{I}={\mathrm{De{}}}^{I} = {\varpi {}}^{I}{}_{K}\wedge
{\mathrm{e{}}}^{K}\\
\end{equation}

The 1915 GR Einstein-Hilbert pure gravity field action density is
the 0-form

\begin{equation}
{\mathrm{L{}}}_{G} = {\epsilon {}}_{\mathrm{IJKL}}{\mathrm{R{}}}^{\mathrm{IJ}}\wedge
{\mathrm{e{}}}^{K}\wedge {\mathrm{e{}}}^{L}\\
\end{equation}
\begin{equation}
S_G =\int L_Gd{}^{4}x.
\end{equation}
\begin{equation}
\frac{{\delta S{}}_{G}}{{\delta e{}}^{I}}=0\\
\end{equation}
is the pre-Feynman action principle. It is the critical point pure
gravity vacuum classical field equation in the absence of matter field
sources, that in the usual tensor notation is
\begin{equation}
{\mathrm{R{}}}_{\mu \nu } = 0.
\end{equation}

Adding the on-mass-shell matter fields of real quanta plus their
off-mass-shell virtual quanta, which contribute to the cosmological
scalar $\Lambda$${}_{\mathrm{zpf\ }} $field, replaces the above
classical vacuum curvature field equation. The vanishing functional
derivative of the {\em total} action $S$ with respect to the 4 GCT
invariant tetrad 1-forms\footnote{The tetrad 1-forms are the
  compensating gauge fields from localizing the global space-time
  universal translation symmetry group acting equally on all matter
  field actions.} i.e:

\begin{equation}
\frac{\delta S}{{\delta e{}}^{I}}=0
\end{equation}
\begin{equation}
S = {\mathrm{S{}}}_{G} + {\mathrm{S{}}}_{M} + {\mathrm{S{}}}_{\mathrm{zpf}}
\end{equation}
\begin{equation}
{\mathrm{R{}}}_{\mu \nu } -\left( \frac{1}{2}R + \Lambda {}_{\mathrm{zpf}}\right)
{\mathrm{g{}}}_{\mu \nu }=-\frac{8\pi G}{{\mathrm{c{}}}^{4}}{\mathrm{T{}}}_{\mu
\nu }
\end{equation}

We can no longer assume the zero torsion field limit ($\varpi^{\mathrm{IJ}}=0$) of the Bianchi
identity of 1915 GR. And so we dream that the final theory will use a more
general connection than that of Levi-Civita.  This more general
connection is induced by locally gauging a more general spacetime
symmetry group (e.g. the Poincare\cite{Kibble}, de Sitter or perhaps the conformal group)
instead of gauging $T(4)$ as is usually done in GR. 

We use ``$;$'' and ``$|$'' to denote covariant differentiation with
respect to the LC connection and the more general connection,
respectively, and so rewrite the divergence of the Einstein equation as:

\begin{equation}
{\mathrm{G{}}}_{\mu \nu }{}^{|\nu }+\frac{\partial \Lambda {}_{\mathrm{zpf}\ \ }}{\partial
{\mathrm{x{}}}^{\nu }}g^{\nu}_{\mu} +\frac{8\pi G}{{\mathrm{c{}}}^{4}}{\mathrm{T{}}}_{\mu
\nu }{}^{|\nu } = 0
\end{equation}

This more general divergence (flow) equation with its additional
stress-energy currents suggests new channels inter-connecting the
incoherent random vacuum zero point fluctuations to the smooth
coherent (generalized) curvature field.  However, it must be
supplemented by a similar torsion field equation that comes from
the vanishing functional derivative of the total action S with respect
to the 6 dynamically independent spin connection $\varpi^{IJ}$ 1-forms.
The  $\varpi^{IJ}$ form the compensating gauge field induced from localizing the
6-parameter homogeneous Lorentz group $SO(1,3)$ that also acts equally
on all matter field actions, i.e.,
\begin{equation}
\frac{\delta S}{{\delta \varpi {}}^{\mathrm{IJ}}}=0
\end{equation}

The extra-dimensional brane M-Theory, when mature, should predict a
renormalization group flow to larger gravity coupling strength in the
short-distance limit.  This, when combined with the world hologram
conjecture, suggests Abdus Salam's 1973 bi-metric f-gravity \cite{Salam} formula

\begin{equation}
G= {\mathrm{G{}}}_{\mathrm{Newton}}( 1 + {\alpha e{}}^{-r/\Delta L})
\end{equation}
\begin{equation*}
\alpha \gg 1
\end{equation*}
has been rediscovered by brane theorists \cite{Esposito} looking at the Calabi-Yau (torsion)
field.

\section{Conclusions}

In this model there is no quantum gravity in the usual sense of
starting with a classical field and quantizing it. Rather, we go the
opposite way in the spirit, though not the letter, of Sakharov's 1967
proposal \cite{Sakharov}. What hitherto was called the classical
gravity field is seen to be really an emergent effective {\em
  macro-quantum coherent} c-number post-inflation vacuum field.

We claim the residual random negative-zero-point-pressure advanced
virtual bosons {\em back-from-the-future} manifest as the
anti-gravitating universally repulsive dark energy. This is because
the future de Sitter horizon for a co-moving observer in our universe
is a Wheeler-Feynman perfect absorber -- an infinite red shift surface
-- just like a black hole event horizon is for a static LNIF observer.

In contrast, we claim the universally attracting dark matter comes
from residual positive-zero-point-pressure virtual fermion-antifermion
pairs. In this picture, looking for real on-mass-shell dark matter
particles in the LHC or in underground WIMP detectors is like looking
for the motion of the Earth through the mechanical aether of Galilean
relativity using Michelson and Morley's Victorian interferometer.

This model pocket universe, relative to our Earth-bound detectors, is
created nonlocally and self-consistently by what Igor Novikov and Kip
Thorne call ``a loop in time'' \cite{Novikov}. Advanced Hawking
radiation is emitted from the future cosmic (``de Sitter'') horizon at
all times and is blue-shifted as it travels into the past. However,
only the advanced radiation emitted at time $t_{trigger} \approx 4$
Gyr arrives back at $t=0$, where it ignites the big bang (see figure 1.1 in \cite{Davis}).

Advanced information (Hawking radiation) flows from the infinite
future proper time ``Omega'' $S^2$ horizon of the observer's world line
(shown in the lower ``conformal time'' ($\tau$) diagram in \cite{Davis} Fig
1.1) down the null cosmic event horizon to its intersection with the
null particle horizon (at $t_{trigger}\approx4$Gyr,
$\tau\approx32$Gyr). It flows back along the particle horizon, winding
up at the initial ``Alpha'' moment of inflation completing what John
Cramer calls a ``transaction''\cite{Cramer}.

Furthermore, looking at Fig 5.1 of
\cite{Davis} we see that the future cosmic de Sitter horizon area $A_c \sim
N(t)$, where $t$ is the proper cosmic Robertson-Walker time (corresponding to
spacelike hypersurfaces of maximal CMB isotropy), rises from ``zero'' (1
Bit) to de Sitter asymptote $N(t\to\inf) \approx 10^{124}$ Bits rather
quickly. The advanced dark energy thermal Hawking radiation reaching
us now backward-through-time along our future light cone is very close
to the asymptotic value. 

The final entropy of our (retro)-causal\cite{Cramer}\cite{Price}
universe is $Nk_B \approx 10^{124}$ Bekenstein bits, as mandated by
the area of the future de Sitter horizon, whereas the initial entropy
of the universe is exactly $k_B = 1$ bit, as mandated by the (Planck)
area of the initial singularity.  

This shows very clearly why the cosmological arrow of time is aligned
with the thermodynamic arrow of time, solving Roger Penrose's main
objection \cite{Penrose} to inflationary cosmology: why the early
universe has relatively low entropy.  We need both retrocausality and
the world hologram principle to properly understand the Arrow of Time
of the Second Law of Thermodynamics.

\section{Background related reading}

The works of John Wheeler and Richard Feynman \cite{Wheeler}, Fred
Hoyle and Jayant Narlikar \cite{Hoyle}, James Woodward \cite{Woodward},
Michael Ibison \cite{Ibison}, Robert Becker \cite{Becker} and John
Cramer \cite{Cramer} on advanced electromagnetic waves, radiation
reaction and vacuum fluctuations, Leonard Susskind \cite{Susskind} and
Jack Ng's work on the world hologram \cite{Ng}, Gennady Shipov on
torsion fields \cite{Shipov}, and finally Hagen Kleinert's work on
singular multi-valued phase transformations as local gauge
transformations \cite{Kleinert} are essential background reading.  We
have also found Chapter 2 of Rovelli's lectures on quantum gravity
\cite{Rovelli} very clear regarding the use of Cartan forms in gravity
theory.

During the preparation of this paper we be became aware of a publication entitled: ``Is
dark energy from cosmic Hawking radiation?''\cite{Lee} that relates $w\!
= -1$ dark energy to observer-dependent Hawking radiation. However,
those authors do not clearly specify which horizon they are referring
to. They omit the (key) notion of retro-causality.  We claim
retrocausality is necessarily implied when the correct cosmic
horizon - the {\em future lightlike} de Sitter horizon -- is
specified. Tamara Davis's thesis \cite{Davis} (especially her figures
1.1 and 5.1) clarifies much of the rampant confusion over this subtle issue.

\end{document}